\documentclass[a4paper,UKenglish]{article}
 
\usepackage{microtype}

\usepackage{bm, amsmath, amssymb, amsthm, amsfonts}

\usepackage{cite}
\usepackage{framed}
\usepackage[framemethod=tikz]{mdframed}
\usepackage{appendix}
\usepackage{graphicx}
\usepackage{color}
\usepackage{varwidth}
\usepackage{wrapfig}
\usepackage{algorithm2e}
\usepackage[noend]{algpseudocode}
\usepackage[textsize=tiny]{todonotes}

\newcommand{\changed}[1]{{{#1}}}

\usepackage[textsize=tiny]{todonotes}

\usepackage{framed}
\usepackage[framemethod=tikz]{mdframed}
\usepackage[bottom]{footmisc}
\usepackage[shortlabels]{enumitem}
\setitemize{noitemsep,topsep=3pt,parsep=3pt,partopsep=3pt}
\usepackage[font=small]{caption}
\usepackage{xspace}

\newtheorem{theorem}{Theorem}[section]
\newtheorem{lemma}[theorem]{Lemma}
\newtheorem{meta-theorem}[theorem]{Meta-Theorem}

\definecolor{darkgreen}{rgb}{0,0.5,0}
\usepackage{hyperref}
\hypersetup{
    unicode=false,          
    colorlinks=true,        
    linkcolor=red,          
    citecolor=darkgreen,        
    filecolor=magenta,      
    urlcolor=cyan           
}
\usepackage[capitalize, nameinlink]{cleveref}
\crefname{theorem}{Theorem}{Theorems}
\Crefname{lemma}{Lemma}{Lemmas}
\Crefname{conjecture}{Conjecture}{Conjectures}

\algnewcommand\algorithmicswitch{\textbf{switch}}
\algnewcommand\algorithmiccase{\textbf{case}}

\algdef{SE}[SWITCH]{Switch}{EndSwitch}[1]{\algorithmicswitch\ #1\ \algorithmicdo}{\algorithmicend\ \algorithmicswitch}%
\algdef{SE}[CASE]{Case}{EndCase}[1]{\algorithmiccase\ #1}{\algorithmicend\ \algorithmiccase}%
\algtext*{EndSwitch}%
\algtext*{EndCase}%

\newcommand{\eps}{\varepsilon}

\newcommand{\poly}{\operatorname{\text{{\rm poly}}}}

\renewcommand{\paragraph}[1]{\vspace{0.15cm}\noindent {\bf #1}:}

\newcommand{\FullOrShort}{short}

\ifthenelse{\equal{\FullOrShort}{full}}{
	
  \newcommand{\fullOnly}[1]{#1}
  \newcommand{\shortOnly}[1]{}
  
  }{

    \newcommand{\fullOnly}[1]{}
    \newcommand{\IncludePictures}[1]{}
   
  }

\begin{document}

\title{Tight Analysis for the 3-Majority Consensus Dynamics}
\author{
Mohsen Ghaffari$^{1}$, 
Johannes Lengler$^2$}
\date{
$^1$ETH Zurich, \texttt{ghaffari@inf.ethz.ch}\\
$^2$ETH Zurich, \texttt{johannes.lengler@inf.ethz.ch }
\\[2ex]
\today
}
\maketitle




%

\maketitle

\begin{abstract}
We present a tight analysis for the well-studied randomized 3-majority dynamics of stabilizing consensus, hence answering the main open question of Becchetti et al. [SODA'16].

\medskip
Consider a distributed system of $n$ nodes, each initially holding an opinion in $\{1, 2, \dots, k\}$. The system should converge to a setting where all (non-corrupted) nodes hold the same opinion. This consensus opinion should be \emph{valid}, meaning that it should be among the initially supported opinions, and the (fast) convergence should happen even in the presence of a malicious adversary who can corrupt a bounded number of nodes per round and in particular modify their opinions.
A well-studied distributed algorithm for this problem is the 3-majority dynamics, which works as follows: per round, each node gathers three opinions --- say by taking its own and two of other nodes sampled at random --- and then it sets its opinion equal to the majority of this set; ties are broken arbitrarily, e.g., towards the node's own opinion.

\medskip
Becchetti et al. [SODA'16] showed that the 3-majority dynamics converges to consensus in $O((k^2\sqrt{\log n} + k\log n)(k+\log n))$ rounds, even in the presence of a limited adversary. We prove that, even with a stronger adversary, the convergence happens within $O(k\log n)$ rounds. This bound is known to be optimal. 
\medskip

\hrule
\medskip
\end{abstract}


\section{Introduction and Related Work}
In this paper, we provide a tight analysis for the convergence time of the well-known $3$-majority dynamics for consensus, as investigated before by Becchetti et al.\cite{becchetti2014simple, becchetti2016stabilizing}. This is a very simple probabilistic process which allows a distributed system to converge to consensus on one of the opinions held by one of the nodes in the system, even in the presence of a byzantine adversary with some limited power. Let us go directly into the dynamics. We refer to \cite{becchetti2016stabilizing} for a nice discussion about the motivations and applications of this dynamics in distributed systems.

\paragraph{The 3-Majority Dynamics} Consider a distributed system of $n$ nodes, each initially holding an opinion in $\{1, 2, \dots, k\}$. We usually assume that $k$ is moderately small\footnote{With some more care, our analysis can be extended to larger values of $k$. We leave describing that extension to the journal version of this work.}, in particular it is at most $k=O(\sqrt{n/\log n})$. Communications happen in the classical $\mathsf{GOSSIP}$ model, with synchronous rounds. Per round, each node $v$  takes three opinions, say by taking its own opinion and pulling the opinion of two other nodes chosen at random. Then, node $v$ updates its opinion to the majority of the set of three opinions it sees, two from the randomly pulled nodes and one of its own.\footnote{\changed{This process slightly deviates from the process in \cite{becchetti2016stabilizing}. While in our case the set of three samples is comprised of the node's own opinion and those of two random neighbors, in \cite{becchetti2016stabilizing}, the set of three samples is formed by considering opinions of three random neighbors (with replacement and including the node itself). We decided against the latter variant for two reasons: first, it seems unnatural for a node $v$ to completely ignore its own opinion for the majority vote, and the former algorithm is certainly no harder to implement. Second, while both variants have the same first-order dynamics, the variances in our version are smaller and allow for a wider range of $k$ and a stronger adversary.  However, we believe that our proof also goes through for the model in \cite{becchetti2016stabilizing}, modulo slight modifications, as we outline at the end of \Cref{subsec:AnalysisOutline}.}} Ties are broken towards the node's own opinion. In other words, any node $v$ that holds opinion $i$ will keep its own opinion, unless both of the nodes that $v$ pulled randomly hold opinion $j\neq i$, in which case $v$ switches to supporting opinion $j$. The process has also been called \emph{two-sample voting}~\cite{cooper2014power} and \emph{two-choices protocol}~\cite{elsasser2016rapid}.


A key desirable property of this simple dynamics is that it converges to a \emph{consensus} setting where all nodes support the same opinion. Indeed, the convergence happens even in the presence of an \emph{$F$-bounded adversary}, who can corrupt $F$ nodes per round, for a reasonably small $F$, modifying their opinions to arbitrary \emph{valid} opinions in $\{1, 2, \dots, k\}$ or even to some \emph{non-valid} opinions not in $\{1, 2, \dots, k\}$. In this case, almost all nodes converge to consensus, modulo those corrupted, and the consensus is on a valid opinion. 
The main question of interest in analyzing this process is to characterize the convergence time to consensus. 

\subsection{Prior Analysis and Ours}
\paragraph{Prior Analysis} Results of Doerr et al.\cite{doerr2011stabilizing} show\footnote{Their analysis was for the $3$-median dynamics, where each node updates its value to the median of its size $3$ set of opinions, but when $k=2$, this dynamics is equivalent to the $3$-majority dynamics.} that in the binary case, where $k=2$, the convergence happens within $O(\log n)$ rounds with high probability\footnote{As standard, \emph{with high probability} indicates a probability of at least $1-1/n^{c}$ for a constant $c\geq 2$.}, even for an $F=O(\sqrt{n})$-bounded adversary. This time complexity for the convergence is tight. More generally, it is known by results of Els\"asser et al.\cite{elsasser2016rapid} that the convergence time is lower bounded by $\Omega(k \log n)$ rounds, even without an adversary.

Becchetti et al.\cite{becchetti2016stabilizing} proved an upper bound of $O((k^2\sqrt{\log n+ k\log n})(k+\log n)$ rounds on convergence time of the $3$-majority dynamics. This convergence happens even despite an $F=O(\sqrt{n}/(k^{2.5}\log n))$-bounded adversary. They left improving this upper bound on the convergence time as an open question. Indeed, see the future work section of \cite[Section 5]{becchetti2016stabilizing}, where Bechetti et al. express their strong belief that the $\Omega(k^3)$ behavior in this convergence-time is not tight, and where they discuss the obstacles in improving the bound using their analysis. 

\medskip
\begin{mdframed}[hidealllines=true, backgroundcolor=gray!20]
\vspace{-3pt}
\paragraph{Our Contribution} We prove that the convergence time is $\Theta(k\log n)$ rounds, with high probability. This matches the aforementioned lower bound of Bechetti et al.\cite{becchetti2014simple} and answers the main open question of Becchetti et al.\cite{becchetti2016stabilizing}. This convergence bound holds even in the presence of a stronger adversary, who is $F=O(\sqrt{n}/k^{1.5})$-bounded. 
\end{mdframed}

\subsection{Other Related Work}
We refer the interested reader to \cite{becchetti2016stabilizing}, which covers all the related work for the $3$-majority dynamics and its motivations, and the connections to the well-studied \emph{consensus} problem in various distributed settings. We here briefly mention only a few directly related work. 

There is another regime of interest for the $3$-majority dynamics, where one assumes that at the beginning one of the opinions is much stronger than the others, and then desires that the convergence happens to this plurality opinion. Generally, the type of technical challenges in analyzing that regime is quite different, as indicated and discussed by Becchetti et al.\cite{becchetti2016stabilizing}. \changed{It is known by results of Els\"asser et al.~\cite{elsasser2016rapid} that the convergence in that regime to the plurality opinion happens within $\Theta(k\log n)$ rounds if $k = O(n^{\eps})$, assuming an initial gap of at least $\Omega(\sqrt{n\log n})$ nodes between the plurality opinion and the others. Similar results were obtained by Bechetti et al.~\cite{becchetti2014simple}.} The heart of the technical challenge in our work, and also that of \cite{becchetti2016stabilizing}, is actually in the regime where there is no such gap at the beginning and all opinions start with the same or almost the same amount of support. 

Two other closely resembling simple probabilistic dynamics for consensus have been studied in the literature. One is the \emph{$3$-state} dynamics, which is used and analyzed for binary consensus where $k=2$, in the population-protocols model where per round exactly one chosen pair of nodes interact\cite{angluin2008simple, perron2009using}. Another related dynamics is the \emph{3-median} dynamics where per round, each node updates its opinion to the median of the sampled set of size three. This is studied by Doerr et al.\cite{doerr2011stabilizing} for general $k$. Though, we note that this dynamic is mainly of interest in settings with no adversary; it is known that this dynamic may converge to an invalid opinion --- i.e. an opinion introduced by adversary and not supported by the non-corrupted nodes initially --- even with a very weak adversary where $F=\poly\log n$. See \cite[Section 2]{becchetti2016stabilizing}.

Finally, we note that over the past decade, there has been extensive interest in developing and analyzing simple probabilistic dynamics for distributed system. In particular, a great number of various simple probabilistic distributed dynamics have been studied in somewhat similar settings, either in the uniform gossip model, the gossip  model on graphs, or the population-protocols model. These are motivated by a wide range of application domains spanning chemical reaction networks, biological settings, social networks, and peer-to-peer networks. It is well beyond the scope of this paper to provide an exhaustive coverage of all of these work. Merely as a random sample of some of the most recent work, we mention \cite{alistarh2015fast, becchetti2015plurality, berenbrink2016efficient, ghaffari2016polylogarithmic, alistarh2017time}.

\section{Analysis Intuition and Outline}
For simplicity, when providing the analysis, we first discard the adversary. This analysis is presented in a format that allows us to easily incorporate the adversary's effect, as we explain in \Cref{app:adversary}, without changing the asymptotic bounds. In particular, the effect of the adversary, who can change the opinion of at most $F=\epsilon \sqrt{n}/k^{1.5}$ nodes per round, for a small constant $\eps>0$, will be weaker than the random deviations that we take into account in our analysis.

\subsection{Intuition Based on a First-Order Analysis} For each $i\in \{1, 2, \dots, k\}$, let $p_i$ denote the fraction of nodes who hold opinion $i$. Then, in one round, we expect the following change: 
$$\mathbb{E}[p^{new}_i] = p_i + (\sum_{j\neq i} p_j) p^2_i - p_i (\sum_{j\neq i} p^2_j).$$ Here, the second term indicates the expected fraction of nodes of opinion $j\neq i$, summed up over all $j$, who sample two nodes of opinion $i$ and thus join opinion $i$. The third term indicates the expected fraction of nodes of opinion $i$ who sample two nodes of another opinion $j\neq i$, summed up over all $j$, and thus migrate to opinion $j$. Using notation $\Sigma_2 = \sum_{j} p_i^2$ and the fact that $\sum_{j} p_j =1$, we now rewrite this expected change in a more convenient format:
$$\mathbb{E}[p^{new}_i] = p_i (1 + p_i - \Sigma_2).$$

Of course the above equality describes only the expected change in $p_i$. The actual change will be not be precisely the same, but it will have some concentration around this mean. However, this concentration will not be sufficiently strong for us, to allow taking all the possible deviations of all the rounds into account in the worst case (e.g., via a union bound). Furthermore, sometimes we actually want to argue that some anti-concentration type phenomena happens and the support of different opinions get some minimal difference, due to the variances. Before these, let us provide some intuition by examining an idealized behavior where we ignore the deviations and assume that the system evolves according to its first-order approximation, that is, $p^{new}_i \leftarrow p_i (1 + p_i - \Sigma_2)$. 

The hardest regime for the analysis is when most opinions are close to the maximum size opinion. Once an opinion is considerably weaker than the plurality opinion, say by a $2$ factor, we will have a much easier time showing that this weak opinion will die out soon. To focus on the core regime of interest, suppose that all the opinions have almost the same size, i.e., where we have $p_i = \Theta(1/k)$ for all $i\in \{1, 2, \dots, k\}$. 

Consider two opinions $i$ and $j$, and without loss of generality suppose that $p_i\geq p_j$. Define $g_{ij} = \frac{p_i - p_j}{p_j}$. That is, $\frac{p_i}{p_j} = 1+g_{ij}$. Then, we have $$1+g^{new}_{ij} = \frac{p^{new}_{i}}{p^{new}_{j}} = \frac{p_{i}}{p_j} \cdot \frac{1+p_i -\Sigma_2}{1+p_j -\Sigma_2} = (1+g_{ij}) \cdot (1+\frac{p_i-p_j}{1+p_j-\Sigma_2}) \geq (1+g_{ij}) \cdot (1+ g_{ij}\cdot \frac{p_j}{2}).$$
This implies that $g_{ij}$ grows like $g^{new}_{ij}\geq {g_{ij}}(1+p_{j}/2)$. Suppose that we start with a minimal non-zero difference between $p_i$ and $p_{j}$, which implies that at the beginning $g_{ij}\geq 1/n$. We note that this initial difference will not be sufficient once we bring back the deviations into our analysis. But for now, this simplistic assumption delivers an instructive intuition. If $g_{ij}\geq 1/n$, so long as $p_j$ does not drop below $\Theta(1/k)$, per round this gap parameter $g_{ij}$ grows by at least a factor of $(1+\Omega(1/k))$. Hence, if that continues for more than $\Theta(k\log n)$ rounds, it would imply $g_{ij}\geq n^3$, which due to the integrability considerations, would mean $p_j=0$. Thus, we conclude that within $O(k\log n)$ rounds, $p_j$ must drop to $o(1/k)$. This means that opinion $j$ is effectively out of the race.

If we could prove some analysis resembling the above for every pair of opinions, we would be done. That is, if one could show that a similar gap shows up between any pair of opinions, i.e. at least one of the two opinions drops below $o(1/k)$, then effectively all but one opinion are out of the race. Once one of the opinions is say a two  factor larger than the others, it will be easy to show that this plurality opinion will reach consensus within $O(k\log n)$ additional rounds. However, analyzing opinions that are very close to each other in size is quite non-trivial, and requires much care due to the likely deviations from the expected behavior.

\subsection{Analysis Outline}\label{subsec:AnalysisOutline}
The above discussions provides some intuition for how the process evolves. We next describe the high-level outline of how we turn this intuition into an analysis. 

Throughout, when talking about a round $t$, we use $p_1$, $p_2$, \dots $p_k$ to denote the fraction of nodes of opinions $1$, $2$, \dots $k$, respectively, at time $t$. 
For the sake of simplicity, we do not express the fact that these fractions change with time $t$ explicitly in our notations. When needed, we will use superscripts to indicate the time to which these parameters correspond. Sometimes, when talking about a given interval of time, we use the superscript $new$ --- for instance by writing $p^{new}_{i}$ --- to indicate the values at the end of the interval.

\paragraph{Super-Weak, Weak, and Strong Opinions} We call an opinion $i$ \emph{super-weak} if its support  $p_i \leq {1}/{(10k)}$. Even summed up over all opinions, the super-weak opinions can take only a ${1}/{10}$ fraction of the nodes. Thus, at least a $9/10$ fraction of the support is on \emph{not-super-weak} opinions. Furthermore, we call an opinion $i$ \emph{weak} in a given round if $p_i \leq p_{max}/5$. Here, $p_{max}=\max_{i} p_i$ in that round. If $p_i \geq p_{max}/5$, we call opinion $i$ a \emph{strong} opinion in that round. 

We will see that super-weak opinions remain super-weak, with high probability. Moreover, weak opinions are also effectively out of the race, because as we shall prove, weak opinions remain weak, with high probability. Moreover, each weak opinions becomes super-weak at some point and thus falls completely out of the race. Our core attention will be on \emph{strong} opinions.  
\begin{lemma}(\textbf{Property (P1)})\label[lemma]{lem:P1} With high probability, super-weak opinions remain super-weak, and weak opinions remain weak.
\end{lemma}

\paragraph{Epochs of the Analysis} 
We break time into \emph{epochs}, each made of an interval of consecutive rounds. This is done in a way that at all times during the $i^{th}$ epoch, the number of opinions that are not super-weak is at most $\lfloor{k (5/6)^{i-1}\rfloor}$.

Let us consider the $i^{th}$ epoch, and let $\kappa = \lfloor{k(5/6)^{i-1}\rfloor}$. Suppose that $\kappa \geq 2$. Notice that at the beginning of this epoch, we have $p_{max}=\max_{i} p_i \geq \frac{0.9}{\kappa}$. This is because there are at most $\kappa$ opinions who are not super-weak, and these opinions should have at least a $9/10$ fraction of nodes in total. If at any time during the epoch, we reach a setting where $p_{max}= \max_{i} p_i \geq \frac{1.5}{\kappa}$, then we say that the \emph{end-of-time} has arrived for this epoch. 

It will be easy to show that once the end-of-time arrives, within at most $O(\kappa \log n)$ additional rounds, only $5\kappa/6$ opinions remain who are not super-weak, and hence, the next epoch begins. The most interesting regime of the analysis is before the \emph{end-of-time}. 

We show that within $O(\kappa \log n)$ rounds from the beginning of the epoch, with high probability, the end-of-time arrives. We discuss this part soon. Once we have that, we can infer that this epoch takes at most $O(\kappa \log n)$ rounds overall, summed up over the period before the end-of-time and the period after that. Hence, thanks to the geometric decay of $\kappa$ throughout different epochs, even summed up over all epochs, the time complexity is $O(k\log n)$. This is until we reach a setting where $\kappa=1$ and only one not-super-weak opinion remains. This single opinion must have at least $9/10$ fraction of the support, as the super-weak opinions can in total amount to at most a $1/10$ fraction of the support. It is easy to see that this majority opinion will get everyone's support within $O(\log n)$ additional rounds, with high probability.

\paragraph{Phases of an Epoch} The most interesting part of the analysis of each epoch is showing that the \emph{end-of-time} arrives soon. Concretely, we prove that starting in an epoch where only $\kappa$ not-super-weak opinions exist, we get to the end-of-time where $p_{max}=\max_{i} p_i \geq \frac{1.5}{\kappa}$, within $O(\kappa \log n)$ rounds. We break this period of time into \emph{phases}. We define each \emph{phase} to consist of $\delta \kappa$ consecutive rounds, for a desirably small constant $\delta>0$. This length is chosen so that the change in the opinions during one phase is small. Let us explain. So long as we are within this epoch and the end-of-time has not arrived, we have $\max_{i} p_i \leq \frac{1.5}{\kappa}$. As such, the expected fraction of nodes that change their opinion during one round is at most $\sum_{j} p^2_j \leq \max_{i} p_i \leq \frac{1.5}{\kappa}$. As we shall see later, a similar upper bound holds also with high probability. Hence, even over all the $\delta \kappa$ rounds of this phase, we expect no more than $3\delta \ll 1$ fraction of nodes to change their opinions. Choosing $\delta$ a small enough constant allows us to think that the change during the phase is relatively small. In some sense, this gives us sufficient \emph{smoothness} during the phase, which allows us to ignore some smaller-order effects.

\paragraph{Seven Key Properties In Analyzing Phases and Epochs} We will establish six key properties, aside from the property (P1) stated above, for the analysis of each epoch. Properties (P2) to (P5) are about the period of the epoch before the end-of-time. Hence, during properties (P2) to (P5), we assume that the end-of-time has not arrived yet and we have $\sum_{i} p_i \leq \frac{1.5}{\kappa}$. \changed{Note that this does not mean that we condition on the number of rounds until the end-of-time arrives. Rather, we condition on the history of the process. This condition $\sum_{i} p_i \leq \frac{1.5}{\kappa}$ is checked at the beginning of the round (in the analysis), and if it is not satisfied, the analysis moves to the end-of-time period. The properties below focus on strong opinions.} Notice that during this time, $p_{\max} \in [\frac{0.9}{\kappa}, \frac{1.5}{\kappa}]$ and thus each strong opinion $i$ has $p_i \in [\frac{0.18}{\kappa}, \frac{1.5}{\kappa}]$. Properties (P6) and (P7) handle how the end-of-time arrives and we exit this epoch. 

For properties (P2) to (P5), we use a key auxiliary definition, which captures the gap between the size of two (strong) opinions. For two strong opinions $p_i$ and $p_j$, we define $g_{ij} = \frac{p_i-p_j}{p_j}$ and $g_{ji} = \frac{p_j-p_i}{p_i}$. Alternatively, we can write $1+g_{ij} = \frac{p_i}{p_j}$. In the most interesting regime of the analysis, $\frac{p_i}{p_j} \approx 1$ and thus, we effectively think of $g_{ij}$ as the second-order term in $\frac{p_i}{p_j}$.

For all the following lemmas, we assume that $n\geq n_0$ for a sufficiently large constant $n_0\geq 2$. 

\begin{lemma} (\textbf{Property (P2)})\label[lemma]{lem:P2} Consider two strong opinions $i$ and $j$. For any desirably large constant $C_1 >0$, by the end of the phase, we have $\max\{g_{ij}, g_{ji}\} \geq \frac{C_1}{\sqrt{n/\kappa}}$, with a probability at least $C_2> 0$, for a constant $C_2$ that depends on constant $C_1$.
\end{lemma}

\begin{lemma} (\textbf{Property (P3)})\label[lemma]{lem:P3} Consider two strong opinions $i$ and $j$, and suppose that at the start of the phase, $p_i\geq p_j$. Furthermore, assume that $g_{ij} = \frac{x}{\sqrt{n/\kappa}}$, for $x\geq C_1$, for a desirably large constant $C_1>0$. Then, we have $\Pr[g^{new}_{ij} \geq (1+\delta/100) \cdot g_{ij}] \geq 1-exp(-C_4x^2)$, where $g^{new}_{ij}$ denotes the gap parameter at the end of the phase, and $C_4$ is a constant that depends on $C_1$.
\end{lemma}

\begin{lemma} (\textbf{Property (P4)}) \label[lemma]{lem:P4} For any two strong opinions $i$ and $j$ and any desirably large constant $C_5\geq 0$, with high probability, in $O(\log n)$ phases, we will have $\max\{g_{ij}, g_{ji}\} \geq \frac{C_5\sqrt{\log n}}{\sqrt{n/\kappa}}$.
\end{lemma}

\begin{lemma} (\textbf{Property (P5)})\label[lemma]{lem:P5} Consider two strong opinions $i$ and $j$, and suppose that at the start of the phase, $p_i\geq p_j$. Furthermore, assume that $g_{ij} \geq \frac{C_5\sqrt{\log n}}{\sqrt{n/\kappa}}$, for a desirably large constant $C_5>0$. Then, by the end of the phase, we have $g^{new}_{ij}\geq g_{ij} \cdot (1+\delta/100)$, with high probability.  
\end{lemma}

\begin{lemma} (\textbf{Property (P6)})\label[lemma]{lem:P6} 
Consider the $i^{th}$ epoch where at most $\kappa=\lfloor k (5/6)^{i-1}\rfloor \geq 2$ not-super-weak opinions exist, at the beginning. With high probability, within $O(\kappa \log n)$ rounds, the end-of-time arrives and we have $p_{max}=\max_{i} p_i \geq \frac{1.5}{\kappa}$.
\end{lemma}

\begin{lemma} (\textbf{Property (P7)})\label[lemma]{lem:P7} Suppose that in the $i^{th}$ epoch, which starts with at most $\kappa = \lfloor k (5/6)^{i-1}\rfloor$ not-super-weak opinions, we have reached the end-of-time and have $\max_{i} p_i \geq \frac{1.5}{\kappa}$. Then, within $O(\kappa \log k)$ additional rounds, with high probability, at most $5\kappa/6$ not-super-weak opinions remain and thus, the next epoch begins.
\end{lemma}

\changed{We still assume that there is no adversary, and add the adversary later in \Cref{app:adversary}. Without adversary, \Cref{lem:P7} implies the desired runtime bound, because we can iteratively apply \Cref{lem:P7} until we reach $\kappa <2$. As explained earlier, since $\kappa$ drops exponentially, this takes in total only $O(k \log n)$ rounds if the $O(\kappa \log k)$ bound holds in each application of \Cref{lem:P7} (which is true with high probability by a union bound). Once $\kappa <2$ is reached, there is at most $\lfloor \kappa \rfloor =1$ opinion left which is not-super-weak. By definition of super-weak, this opinion will have at least $9/10$ of the support. Afterwards, it is easy to see that in each further round, with high probability the number of nodes with a different opinion decreases by at least a factor $1/2$ until there are only a constant number of nodes with different opinions left. Those are swallowed by the majority within a constant number of additional rounds, with high probability. Altogether, with high probability the algorithm terminates after $O(k\log n)$ rounds, as desired.
}

\paragraph{Proof Styles} Let us discuss the type of arguments that goes into proving these seven properties. Properties (P1) and (P7) will be established using standard concentration arguments. Properties (P2), (P3), and (P5) will require the most care and effort. Proofs of (P3) and (P5) are very close, and they are stated separately merely for convenience. These three properties are established by closely analyzing the behavior during one phase, and showing rather tight anti-concentration or concentrations for the process. Especially in the concentration case, we will need the per-round analysis to be sharp modulo the third-order term. Property (P4) is much less complex, particularly because it does not deal with the specifics of our dynamic directly; it will be established using arguments that are somewhat standard for the so-called \emph{explosive processes}, where one wishes to prove that a dynamics polarizes to one side or the other. Finally, property (P6) is a rather straightforward consequence of properties (P4) and (P5).

\paragraph{Adaptation for 3-Majority with 3 Random Nodes} \changed{The paper \cite{becchetti2016stabilizing} studied a slightly different specification of the 3-majority consensus dynamics, where each node $v$ picks three nodes uniformly at random, and takes the majority of these opinions (breaking ties randomly). This process has the same first order dynamics, but the variances are higher. However, in our analysis we mostly need to compare variances \emph{with each other}, so essentially the proof still goes through. Here we list only the most important changes, skipping the details.
\begin{itemize}
\item The variances increase by a factor of $\kappa$. In particular, the variance of the number of blue vertices (of a strong opinion) is $\Theta(n/\kappa)$ instead of $\Theta(n/\kappa^2)$. So for example, we expect the number of light blue vertices after the first round of an epoch to be of order $\sqrt{n/\kappa}$ instead of $\sqrt{n/\kappa^2}$. Similarly, for each phase the variance of the number of blue vertices is $\Theta(n)$ instead of $\Theta(n/\kappa)$, and we expect the deviations to be of order $\sqrt{n}$ instead $\sqrt{n/\kappa}$. \\
More formally, the factors $1/\sqrt{n/k}$ in Lemmas \ref{lem:P2}, \ref{lem:P3}, \ref{lem:P4}, and \ref{lem:P5} need to be replaced by $1/\sqrt{n}$. 
\item For \Cref{lem:P1} we need that weak opinions remain weak. Thus we need that if $g_{ij} = \Omega(1)$ is sufficiently large, the expected gain of $g_{ij}$ in each round dominates the deviation. The gain is $\Omega(1/\kappa)$, while the standard deviation of $g_{ij}$ is $O(\sqrt{\kappa/n})$ (instead of $O(1/\sqrt{n})$). Thus we need $1/\kappa \gg \sqrt{\kappa/n}$, or $\kappa \ll n^{1/3}$, where an additional polylogarithmic factor is needed to ensure high probability. 
\item For the proof we need to adapt our definition of extra-light blue nodes. We say that a light-blue node $u$ recruits an extra-light blue node $v$ if $v$ changes its opinion to blue in a round in which it saw $u$ \emph{and another blue node}. If a node $u$ changes an opinion due to a light-blue node in a tie, then we color $u$ light-blue. In this way, the number of extra-light blue nodes is the same as in our analysis, and can be treated analogously. 
\item The treatment of light-blue nodes becomes slightly more complicated: let $x_{i-1}$ and $x_i$ be the number of light blue vertices in round $i-1$ and $i$, respectively. Then in our proof we use that $x_i$ equals $x_{i-1}$ plus the deviation from mean of the number of clear blue nodes in round $i$ (which may be positive or negative). For the model from \cite{becchetti2016stabilizing}, the term $x_{i-1}$ must yet be replaced by a binomially distributed random variable $Z$, so the variations of $Z$ cause additional variance in $x_i$. However, $Z$ is binomially distributed with expectation $x_{i-1}$, so it can be handled with Azuma's inequality. Since the variance of $Z$ is strictly dominated by the variance in the clear blue vertices, the additional terms are negligible. 
\end{itemize}
 }


\section{Details of the Analysis: Proving the Seven Properties}
The proofs of properties (P1), (P4), (P6), and (P7) are simpler and are deferred to \Cref{app:details}. Here, we discuss the proof of property (P5), which is one of the main ingredients of the analysis. Property (P5) is itself a special case of property (P3). We will present the proof in a format that can be easily extended to property (P3). Property (P2) also fits the same framework, and is in fact somewhat simpler. Proofs of properties (P3) and (P2) appear later in \Cref{subsec:P2-3}. 


%
\subsection{Property (P5)}
In property (P5), we focus on one phase, which is made of $\delta \kappa$ rounds, for a constant $\delta>0$ that will be chosen desirably small. Moreover, we assume that during all rounds of this phase, we have $p_{max} = \max_{i} p_i \leq  1.5/\kappa$. Once $p_{max}$ exceeds $1.5/\kappa$, the end-of-time arrives for this epoch, and we invoke \Cref{lem:P7} to infer that within $O(\kappa \log k)$ additional rounds, we move to the next epoch.

\medskip
\noindent\textbf{\Cref{lem:P5}} (\textbf{Property (P5)})\emph{Consider two strong opinions $i$ and $j$, and suppose that at the start of the phase, $p_i\geq p_j$. Furthermore, assume that $g_{ij} \geq \frac{C_5\sqrt{\log n}}{\sqrt{n/\kappa}}$, for a desirably large constant $C_5>0$. Then, by the end of the phase, we have $g^{new}_{ij}\geq g_{ij} \cdot (1+\delta/100)$, with high probability.} 
\medskip

\begin{proof}
We begin the proof by some intuitive discussions, mainly based on expectations. Then, we present a coloring scheme that allows us to track how far off the process goes from these expectations. Then, we bound the components of this coloring and thus show limits on how far we may be from the expectations. At the end, we put the expectation-based analysis together with the bounds on the deviations to complete the proof.

\paragraph{Intuitive Discussions for the Analysis} To prove the lemma, we focus on only opinions $i$ and $j$ and we closely examine the changes of their support during the $\delta \kappa$ rounds of the phase. We will not monitor the changes in the other opinions, except that we know that at all times during this phase, we have $p_{max} \leq  1.5/\kappa$. This is because otherwise the end-of-time arrives and we soon move to the next epoch. Focusing on two strong opinions $i$ and $j$, let us call nodes who support opinion $i$ \emph{blue} and nodes who support opinion $j$ \emph{red}. 

Let us starting with the first round of the phase. During this round, we have the following expected behavior. The two populations of \emph{blue} and \emph{red} will have only a small change. We expect $p_i \Sigma_2$ fraction of nodes to migrate out of the blue region, and $p^2_i$ fraction of nodes to migrate in to the blue region. Similarly, we expect $p_j \Sigma_2$ fraction of nodes to migrate out of the red region, and $p^2_j$ fraction of nodes to migrate in to the red region. Notice that all of these terms are changes of the order $\Theta(1/\kappa^2)$, which occur on blue and red populations. Note that since $i$ and $j$ are strong, we have $p_i \in [\frac{0.18}{\kappa}, \frac{1.5}{\kappa}]$ and $p_j \in [\frac{0.18}{\kappa}, \frac{1.5}{\kappa}]$. Thus, each of blue and red populations have size in the order of $\Theta(1/\kappa)$. Therefore, in relative terms, the changes are in the order of $\pm\Theta(1/\kappa)$.

Our main analysis focus is on $g_{ij}$, which is the second-order term in the ratio $\frac{p_i}{p_j}$. This is because $\frac{p_i}{p_j} = 1+ g_{ij}$. In this regard, at least in expectation, the outwards migrations are linearly proportional to the current population (with a $\Sigma_2$ factor) and thus, the outwards migrations will not skew the ratio $\frac{p_i}{p_j}$. On the other hand, the inwards migrations favor the blue region slightly, as the blue region is slightly bigger. Hence, we expect that the blue region grows slightly faster. Concretely, in terms of expectations, we expect that by the end of round, $p^{new}_i= p_i (1 + p_i - \Sigma_2)$ and $p^{new}_j= p_j (1 + p_j - \Sigma_2)$. If the changes are sharply concentrated around these means, in terms of the ratio $\frac{p_i}{p_j}$, this is a growth of $\frac{(1 + p_i - \Sigma_2)}{(1 + p_j - \Sigma_2)} \geq 1+ \frac{p_i-p_j}{2} \geq 1+\frac{0.18 g_{ij}}{2\kappa}$ factor. Since $\frac{p_i}{p_j} = 1+ g_{ij}$, that would be a $1+\frac{1}{20\kappa}$ growth in $g_{ij}$. If a similar behavior continues over the next $\delta \kappa$ rounds, during this phase, $g_{ij}$ sees a growth by a factor of $(1+\frac{1}{20\kappa})^{\delta \kappa} \geq 1+\frac{\delta}{30}$. Of course, the whole challenge is that the concentrations are not strong enough to let us say that despite the likely deviations, such a growth still occurs per round.

Let us check the deviations in the blue population. In the above analysis based on expectations, we said that we expect $p_i \Sigma_2 n$ nodes to migrate out of the blue region and $p^2_i n$ nodes to migrate in to the blue region. Notice that both of these are less than $(1.5/\kappa)^2 n$ nodes, because $\Sigma_2 \leq p_{max}\leq \frac{1.5}{\kappa}$. Clearly, we will have some deviation around these mean. In particular, Chernoff bound tell us that the actual number of nodes that move in/out can be off from its expectation by at most an additive $\pm C_6 \sqrt{n \log n/\kappa^2}$, for some constant $C_6>0$. In relative terms compared to $p_i$, this is no more than a $\pm C_7\sqrt{{\log n}/{ n}}$ additive deviation for the change in the value of $p_i$, because $p_i \in [\frac{0.18}{\kappa}, \frac{1.5}{\kappa}]$. In other words, with high probability, $p^{new}_i \in \mathbb{E}[p^{new}_i] \pm p_i \cdot C_7\sqrt{{\log n}/{ n}}.$ That is, for some constant $C_8>0$, we can say that with high probability, $$p^{new}_i \in p_i (1 + p_i - \Sigma_2) \cdot (1\pm C_8\sqrt{\frac{\log n}{ n}}).$$ 

This deviation term $(1\pm C_8\sqrt{{\log n}/{ n}})$ may look tolerable for one round. However, we cannot afford to take it into account in the worst case in each round, as then over the whole phase, the deviation from the expected behavior could be a factor of $(1\pm C_8\sqrt{{\log n}/{ n}})^{\delta \kappa} \approx (1\pm C_8\delta \kappa \sqrt{{\log n}/{ n}})$. Notice that we are working on a parameter $\frac{p_i}{p_j} = 1+g_{ij}$, which itself is quite close to $1$; it can be as small as $1+\frac{C_5\sqrt{\log n}}{\sqrt{n/\kappa}}$. The analysis is mainly on the second order term of the change of $\frac{p_i}{p_j}$ during the phase. Hence, even though we may be able to afford deviation factors up to $\big(1\pm O(\sqrt{\kappa}) \cdot \sqrt{{\log n}/{ n}}\big)$, we certainly cannot afford the $\big(1\pm C_8\delta \kappa \sqrt{{\log n}/{ n}}\big)$ deviation term that would come from worst-case analysis per round. Notice that the difference is roughly an $\sqrt{k}$ factor in the second-order term. This is what we need to put up a fight for! It all boils down to doing the per-round analysis of $\frac{p_i}{p_j}$ in a manner that is tight to within the third-order term. 

The saving grace is that the worst-case per round additive deviations are essentially independent, modulo a smaller effect that we can control. Intuitively, there is no reason that all the $\delta\kappa$ of these deviations terms should go in the same direction, as the above analysis assumed. We will provide an analysis that formalizes this. Very roughly speaking, we effectively show that the overall deviation in terms of the number of nodes is approximately the summation of $\delta\kappa$ zero-mean Gaussians, each with variance at most $O(n(1.5/\kappa)^2)$. This summation is itself a Gaussian of zero-mean and a $\delta \kappa$ factor higher variance. Notice that this is in term of the number of nodes, and not their fraction. When expressed as a fraction of nodes and also relative to $p_i$, this is no more than a deviation factor of $\big(1 \pm O(\sqrt{\kappa} \sqrt{\log n/n})\big)$, with high probability. This deviation is within our tolerable range. 

\paragraph{A Coloring Scheme to Track Deviations in One Phase} To formalize the above intuition and track the deviations from the expected behavior, we use a certain coloring of the nodes. For instance, nodes of opinion $i$ will be colored \emph{blue}, \emph{light blue}, or \emph{extra-light blue}. In particular, the majority of supporters of opinion $i$ will be \emph{blue} nodes, a minority will be \emph{light blue}, and even a much smaller minority will be \emph{extra-light blue}. We will need to follow the population of \emph{blue} nodes sharply, and the population of \emph{light blues} up to a constant factor, but those of \emph{extra-light blue}, we can afford to be much more coarse and use only a simple upper. We will perform a similar coloring for nodes of opinion $j$, coloring them with \emph{red}, \emph{light red}, or \emph{extra-light red}. 

Let us focus on the coloring scheme for nodes of opinion $i$. We start with the very first round of the phase. As mentioned before, we expect the fraction of nodes of opinion $i$ to go from $p_i$ to $p_i + p^2_i - p_i\Sigma_2$. At the end of the round, we will color exactly $p_i + p^2_i - p_i\Sigma_2$ fraction of nodes \emph{blue}. Of course this may be less than or more than the actual number of nodes of opinion $i$, due to the deviations. If we actually have more nodes of opinion $i$ than we colored blue, we will color the left over nodes as \emph{positive-charge light blue} nodes. If we have less nodes in opinion $i$ than we colored blue, then we color a number of those we colored blue equal to the excess as \emph{negative-charge light blue} nodes. Hence, as of now, the supporters of opinion $i$ have two colorings: a \emph{clear blue} color, the number of which is exactly $p_i + p^2_i - p_i\Sigma_2$, and a minority of nodes who have color \emph{light blue}. These light blue nodes may have a positive or negative charge. These charges indicate whether we are above or below the expectation. In the case of negative charge, a node may have both a clear blue color and a negative-charge light blue color. Still, when we talk of nodes with color clear blue, it includes these and it is a population of size exactly equal to $p_i + p^2_i - p_i\Sigma_2$ fraction of nodes. 

We briefly comment that the light blues are indeed a minority. By the Central Limit Theorem, as $n \rightarrow \infty$, the number of light blues in this first round (taking into account their charge) is distributed according to a zero-mean Gaussian. The variance of this Gaussian is no more than the expected number of move in/outs, which itself is at most $\max\{p^2_i, p_i\Sigma_2\} \leq p^2_{\max}$. Hence, with high probability, they are a minority. We will later present a close accounting of the number of light blue nodes, throughout the phase.

We next examine how this coloring evolves during this phase, from one round to the next. This will also be the place where we introduce \emph{extra-light} blue nodes. Without loss of generality, and for the sake of simplicity, let us assume that in this round, we have a body of clear blue nodes, and a minority of light blue nodes of positive charge. The case with negative charges would be similar, just in the opposite direction. 

We will treat clear blue nodes as the main body of nodes of opinion $i$. This means we will essentially neglect the effects of light blue nodes in attracting nodes of other opinions to opinion $i$. Let us examine that closely. Consider a node $v$ of an opinion $\ell\neq i$ that joins opinion $i$, because $v$ sampled two nodes of opinion $i$. It is possible that one or two of these sampled opinion-$i$ nodes were \emph{light blue}. Let us say $u$ is a light blue node who was involved in the recruitment of node $v$ to opinion $i$. In that case, we say node $v$ is a hiring of the light blue node $u$. In that case, we color node $v$ \emph{extra-light blue}. Moreover, node $v$ will remain \emph{extra-light} permanently in our coloring scheme, when analyzing this phase. The reason that we can afford to have this permanent coloring is that, as we will prove, the fraction of such extra-light nodes is quite small, small enough to allow us to almost ignore them, except for using some coarse upper bound. Let us provide an intuitive argument for now, the formal argument will be presented later. 

Intuitively, a light blue node $u$ of opinion $i$ is expected to cause a hiring of at most $2p_{max}\leq 3/\kappa$ many other nodes per round. These hired nodes would become \emph{permanent extra-light} hirings of $u$. Even over all the $\delta \kappa$ rounds, this is an expected hiring of at most $3\delta \kappa/\kappa =3\delta\ll 1$ extra-light blue nodes for node $u$. These \emph{extra-light} blue nodes, who were hired by $u$, can have hirings of their own; any node hired by an extra-light blue is also colored \emph{extra-light} blue and remains extra-light permanently during this phase. Notice that despite this possible tree like growth of hirings to opinion $i$ rooted in the light-blue node $u$, still the expected size of this whole hiring tree rooted in $u$ is small. This is because, the growth of this tree is probabilistically dominated by a Galton-Watson branching process where each node gives birth to an expected of no no more than $3\delta \ll 1$ children. We will be able to conclude that overall this population of extra-light blue nodes is no more than a $10\delta \ll 1$ factor of the light blue nodes, which were created directly due to the deviations. We will come back to formalizing this later. For now, let us ignore these extra-light hirings of light blues, and focus on the light blue nodes themselves.

So far we have only described the set of light blue nodes for the first round of the phase. Now we give the definition for an arbitrary round $t$ during the phase.
Suppose that at the beginning of round $t$, the fraction of the clear blue nodes is $q_i$. If we assume that only clear blue nodes support opinion $i$, we expect this support to go from $q_i$ to $q_i + q^2_i - q_i\Sigma_2$. This will be our setting point of the expectation, in defining the clear blue nodes, that is, we color exactly $q_i + q^2_i - q_i\Sigma_2$ nodes clear blue. This effectively ignores the possible (positive or negative) hirings due to light or extra-light blue nodes. Again, due to the deviations, there might be slightly more or less nodes that end up in opinion $i$, even without switching to or out of opinion $i$ because of meeting the (positive or negative charge) light blue nodes. We will color nodes so that this much of deviation is put in the light-blue nodes, of positive or negative charge, as needed. Furthermore, we will always simplify the charges so that at any time we either only have positive charge light blues, or negative charge light blues. That is, for instance, if right now we have a body of positive charge light blue nodes but the deviation makes us fall below the expectation $q_i + q^2_i - q_i\Sigma_2$, we first cancel enough of the positive light blue nodes, and then if necessary, add sufficient number of negative charge light blues.

\paragraph{Bounding Light Blue Nodes} 
In the first round, the number of light blues is simply the deviation of the move out/in of the initial blue nodes from the expectation. The expected number of nodes that move in/out of blue in one round is no more than $n(1.5/\kappa)^2$. Thus, by the Central Limit Theorem, as $n\rightarrow \infty$, the deviation of these moves from its mean is well-approximated by a zero-mean Gaussian distribution with variance no more than $n(1.5/\kappa)^2$. In other words, the number of light-blue nodes at the end of the first round, including their charge, has a zero-mean Gaussian distribution with variance no more than $n(1.5/\kappa)^2$. 

In the second round, we again have some new deviation from the expectation. As a consequence, the number of light blue nodes changes according to adding a zero-mean Gaussian with variance no more than $n(1.5/\kappa)^2$. Crucially, by our definition of light blues nodes, this change is \emph{independent} of the number of light-blue nodes after the first round. Furthermore, this accounting ignores the extra-light blue nodes, which we will examine later. Similarly, in each next round, the number of light-blue nodes changes according to adding a zero-mean Gaussian with variance no more than $n(1.5/\kappa)^2$, which is independent of the previous rounds.

Recall that the summation of a number of independent random variables, each distributed according to a zero-mean Gaussian, is a random variable that is distributed according to a zero-mean Gaussian itself, with a variance equal to the summation of the variances.\footnote{\changed{Note that it does not matter that we do not know for sure how many variables there are, as the phase might end prematurely because of the end-of-time. If that happens then we just fill up with dummy variables.}} Hence, at the $r^{th}$ round of the phase, the number of light-blue nodes (and their charge) has a zero-mean Gaussian distribution with variance no more than $rn(1.5/\kappa)^2$. In particular, even in the last round $r=\delta\kappa$, the variance is no more than $\delta n\kappa(1.5/\kappa)^2 < 3n/\kappa$. That is, in the last round, the number of light-blue nodes is no more than $C_6\sqrt{n\log n/\kappa}$, with high probability. This number of nodes translates to $\pm C_6\sqrt{\log n/ (\kappa n)}$ fraction of nodes. As $i$ is a strong opinion and thus we have $p_i =\Theta(1/\kappa)$, in relative terms compared to $p_i$, this is a deviation factor of $(1\pm C_7\sqrt{\kappa\log n/n})$ from the expectation, for some constant $C_7>0$.

\paragraph{Bounding Extra-Light Blue Nodes} 
We now bound the number of extra-light blue nodes. Consider a light blue node $u$ in a round $r$ of this phase. We count the number of nodes who join opinion $i$ because of direct or indirect chains of meetings with two opinion-$i$ nodes, that end in the light blue node $u$ in round $r$. If a node $v$ joined opinion $i$ in round $r$, because of meeting two nodes of opinion $i$ one of which was $u$, then we consider $v$ as recruited by node $u$ in round $r$. Moreover, if later on, any node switches to opinion $i$ because of meeting node $v$, or one of the recruits of $v$, those are counted as recruitment of $v$ and thus, indirectly, as recruitment of $u$. Therefore, they are also taken into account when we examine the effect of $u$ being a light blue node in round $r$. 

In round $r$, the number of direct recruitment of $u$ to opinion $i$ is a Binomial distribution with expectation no more than $n \cdot \frac{1}{n} \cdot p_{i} \leq p_{max} \leq \frac{1.5}{\kappa}$. Each recruited node $v$, which becomes an extra-light node, may have its own direct recruitment. Over all the rounds, that is a binomial distribution with expectation no more than $\delta \kappa p_{max} \leq 3\delta \ll 1$. Similarly, any recruited extra-light node may recruit further nodes, according to a binomial distribution with expectation no more than $\delta \kappa p_{max} \leq 3\delta \ll 1$. Hence, the overall number of extra-light blues created because of $u$ being a light-blue node in round $r$ is at most the size of a simple Galton-Watson branching process\cite{watson1875probability}. In this process, first the root gives birth to a Binomially distributed number of children with expectation at most $p_{max}$, and from that point on, each node gives birth to a Binomially distributed number of children with expectation at most $3\delta \ll 1$. It is simple and well-known that in this regime of each node creating strictly less than one off-spring in expectation, the process dies out and moreover, the size of the whole tree has an exponentially decaying probability tail. Concretely, in our case, the probability that the size of the tree exceeds $3t p_{\max}$ decays exponentially with $t$. Therefore, since the summation of random variables with exponentially decaying tail has a Chernoff-like concentration (see e.g., \cite[Lemma 7]{doerr2011stabilizing}), if we let $s^r$ be the fraction of light blue nodes in round $r$, we can conclude that with high probability, the fraction of extra-light blue nodes is no more than $\sum_r |{s^r}| \cdot 4 p_{max} + O(\log n)$. Now, we bound the summation $\sum_r |{s^r}|$. In particular, we bound $\max_{r} |{s^r}|$ using Etemadi's inequality\cite{etemadi1985some}.

\begin{theorem}[Etemadi's Inequality \cite{etemadi1985some}] Let $X_N$ for $N=1,2, 3, \dots$ be a sequence of independent random variables, though not necessarily having identical distributions. For each $L$, define $S_L = \sum_{\ell=1}^{L} X_{\ell}$. For all $x>0$, we have $\Pr\big[\max_{1\leq \ell \leq N} |S_\ell| \geq x\big] \leq 3 \max_{1\leq \ell \leq N} \Pr\big[|S_\ell|\geq x/3\big].$
\end{theorem}
Notice that the fraction of light-blue nodes $s^r$ in round $r$ is itself the summation of $r$ \changed{independent} zero-mean Gaussians, each with variance at most $n(1.5/\kappa)^2\leq 3n/\kappa^2$. In the context of Etemadi's inequality, think of each of these zero-mean Gaussians as one of the summands $X_\ell$. Therefore, Etemadi's inequality shows that $\Pr[\max_{r} |{s^r}| \geq x]$ is no more than 3 times the probability that a zero-mean Gaussian with variance at most $\delta k \cdot 3n/\kappa^2$ exceeds $x/3$. Thus, w.h.p., we have $|{s^r}| \leq 3C_6\sqrt{n\log n/\kappa}$ for all rounds $r$. Hence $\sum_r |{s^r}| \leq 3C_6\delta \kappa \sqrt{n\log n/\kappa} \leq C_{10}\delta \sqrt{\kappa n\log n}$. Therefore, the total number of extra-light blues is with high probability no more than $4p_{max} \cdot C_{10}\delta p_{max} \sqrt{\kappa n\log n} + O(\log n) \ll C_{11}\delta\sqrt{n\log n/\kappa}$, for some constant $C_{11}$. By choosing the constant $\delta$ sufficiently small, we can make this desirably smaller than our bound of $C_6\sqrt{n\log n/\kappa}$ on the number of light-blue nodes. This allows us to treat extra-light blues as negligible compared to the blue nodes. 

\paragraph{Putting Things Together} In the above, we provided arguments that bound the number of light blue and extra-light blue nodes. We can now say that the number of nodes of opinion $i$ at the end of the phase is within a factor of $(1\pm C_{12}\sqrt{\kappa\log n/n})$ of its expectation, for some constant $C_{12}>0$. This expectation is captured by the number of the clear blue nodes. To finish the analysis, we now go back to analyzing this expectation. In the first round, the number of blue nodes is set to $p_i(1+p_i-\Sigma_2)$ and the number of red nodes is set to $p_j(1+p_j-\Sigma_2)$. This is a growth of $\frac{(1+p_i-\Sigma_2)}{(1+p_j-\Sigma_2)}\geq 1+\frac{p_i-p_j}{1+p_j-\Sigma_2} \geq 1+\frac{p_i-p_j}{2} $ factor in the ratio $\frac{p_i}{p_j}$. Similarly, per round, the expectations (captured by clear blue nodes) indicate a growth of at least $1+\frac{p^{t}_i-p^{t}_j}{2}$, where $p^t_i$ and $p^t_j$ indicate the fraction of clear blue and clear red nodes of round $t$. Since these are expectations, it is easy to verify that for all $t$, we have $1+\frac{p^{t}_i-p^{t}_j}{2} \geq 1+\frac{p_i-p_j}{2}$. Hence, at least in terms of the clear blue populations, we expect $\frac{p_i}{p_j}$ to grow by a factor of at least $(1+\frac{p_i-p_j}{2})^{\delta \kappa} \geq 1+\delta g_{ij}/30$. As argued above, each of the two actual fractions may be up to a factor of $(1\pm C_{12}\sqrt{\kappa\log n/n})$ off from these expectations. Hence we can conclude that with high probability, by the end of the phase, we have
$$1+g^{new}_{ij} \geq (1+g_{ij})(1+\frac{\delta}{30} g_{ij}) \cdot (1- \frac{3C_{12} \sqrt{\log n}}{\sqrt{n/\kappa}}).$$ Since we started with the assumption that $g_{ij} \geq \frac{C_5\sqrt{\log n}}{\sqrt{n/\kappa}}$, for a desirably large constant $C_5>0$, we can assume that $3C_{12} \leq \delta C_5/60$, thus allowing us to infer that $g^{new}_{ij} \geq g_{ij}(1+\delta/100).$
\end{proof}

\subsection{Properties (P2) and (P3)}\label{subsec:P2-3}
\medskip
\noindent\textbf{\Cref{lem:P3}} (\textbf{Property (P3)})\emph{Consider two strong opinions $i$ and $j$, and suppose that at the start of the phase, $p_i\geq p_j$. Furthermore, assume that $g_{ij} = \frac{x}{\sqrt{n/\kappa}}$, for $x\geq C_1$, for a desirably large constant $C_1>0$. Then, we have $\Pr[g^{new}_{ij} \geq (1+\delta/100) \cdot g_{ij}] \geq 1-exp(-C_4x^2)$, where $g^{new}_{ij}$ denotes the gap parameter at the end of the phase.}
\medskip

\begin{proof}[Proof Sketch]
As stated before, the proof is quite close to that of \Cref{lem:P5}. We only mention the differences. In proving \Cref{lem:P5}, we needed a high probability guarantee. We thus said that the number of light-blue nodes at the end of the phase, which has a zero-mean Gaussian distribution with variance at most $3n/\kappa$, cannot be more than $\pm C_6\sqrt{n\log n/\kappa}$, with high probability. That is, in terms of fractions, at most $\pm C_6 \sqrt{\log n/(n\kappa)}$. Here, we cannot afford to throw in this $\sqrt{\log n}$ factor for a high probability guarantee. However, just using the definition of a Gaussian, we still can say that the probability that the number exceeds $\pm C_7x/\sqrt{\kappa n}$ is no more than $exp(-C_{13}x^2)$ for some constant $C_{13}>0$. Similarly, using Etemadi's inequality as done before, we get that the probability that the number of light blue nodes ever exceeds $\pm 3C_7x/\sqrt{\kappa n}$ is no more than $exp(-C_{14}x^2)$. This lets us bound the number of extra-light blue nodes, similar to above, by even a smaller order term. Therefore, we conclude that with probability at least $1-exp(-C_4 x^2)$, the deviation factors are bounded by $(1\pm C_{15}x/\sqrt{n/\kappa})$. Hence, with probability at least $1-exp(-C_4 x^2)$, we have $1+g^{new}_{ij} \geq (1+g_{ij})(1+\frac{\delta}{30} g_{ij}) \cdot (1- 3C_{15}x/\sqrt{n/\kappa})$. Since we assume that $C_1>0$ is a desirably large constant, which can be chosen to be larger than say $100 C_{15}/\delta$, we can conclude that with probability at least $1-exp(-C_4x^2)$, we have 
$g^{new}_{ij} \geq (1+\delta/100) g_{ij}$.
\end{proof}

\medskip
\noindent \textbf{\Cref{lem:P2}} (\textbf{Property (P2)})\emph{Consider two strong opinions $i$ and $j$. For any desirably large constant $C_1 >0$, by the end of the phase, we have $\max\{g_{ij}, g_{ji}\} \geq \frac{C_1}{\sqrt{n/\kappa}}$, with a probability at least $C_2> 0$, for a constant $C_2$ that depends on constant $C_1$.}
\medskip

\begin{proof} Suppose that at the beginning of the phase, we have $p_i=p_j$, that is $g_{ij}=0$. We argue that with probability at least $C_2$, for some constant $C_2$ depending on constant $C_1$, by the end of the phase, we will have $g_{ij} \geq \frac{C_1}{\sqrt{n/\kappa}}$. If at the beginning there was a gap, say $p_i>p_j$, that just makes it more likely that $g_{ij} \geq \frac{C_1}{\sqrt{n/\kappa}}$. That step can be formalized using a standard stochastic domination argument. We thus focus on the core case where at the start we have $p_i=p_j$. 

Notice that since at the beginning $p_i=p_j$, the population of clear blue nodes and clear red nodes are the same at the beginning, and they will remain the same throughout the phase. This is simply because the clear colors are always set according to the expectations based on the current clear colored nodes. However, the light blue nodes and light red nodes are created due to the deviations, and they can and will likely differ from each other. 

As argued above, at the end of phase, the number of light-blue nodes, including their charge, is a zero-mean Gaussian. The variance of this Gaussian is at least $(\delta \kappa n(0.18/\kappa)^2) \geq \delta n/(100\kappa)$. This is because per round the deviation is a zero-mean Gaussian with variance at least $n(0.18/\kappa)^2$, as the expected number of out moves is no less than $n p^2_{i}$. Furthermore, the absolute number of extra-light blue nodes is upper bounded, with high probability, 
by $5\delta\cdot\max_{r=1}^{\delta \kappa} |{s^r}|$. Here, $s^{r}$ denotes the number of light-blue nodes in round $r$. 

Consider a desirably large constant $C_{17}>0$, which is chosen sufficiently large as a function of the given constant $C_1$. We are interested in the event that two things happen: ($\mathcal{E}_1$) the number of light-blue nodes, including their charge, exceeds $C_{17} \sqrt{n/\kappa}$, while ($\mathcal{E}_2$) the absolute number of extra-light blue nodes is below $C_{17}\sqrt{n/\kappa}/2$. We argue that there is a constant probability that both of these events happen together.

The probability of the first event ($\mathcal{E}_1$) is some constant $C_{18}=\frac{1}{2}(1 -erf(\frac{C_{17} \sqrt{n/\kappa}}{\sqrt{\delta n/(50\kappa)}})) = \frac{1}{2} (1- erf(\frac{C_{17}}{\sqrt{\delta/50}}))$. Here, $erf()$ is the Gauss error function, which appears in the Cumulative Density Function of a Gaussian distribution. We note that as a function of $x>0$, as $x\rightarrow \infty$, we have $erf(x)\rightarrow 1- e^{-x^2}/2$. Thus, constant $C_{18}$ is quite small, in fact we have $C_{18}\approx \frac{1}{4}exp(-(\frac{C_{17}}{\sqrt{\delta/50}})^2)$. But it is still a constant. On the other hand, by Etemadi's inequality, the probability that the second event ($\mathcal{E}_2$) does not happen is a much smaller constant $C_{19} = \frac{1}{2} \cdot 3 (1- erf(\frac{C_{17}/6 \sqrt{n/\kappa}}{5\delta \sqrt{\delta n/(50\kappa)}}) = \frac{3}{2} \cdot (1- erf(\frac{C_{17}}{\sqrt{\delta/50}} \cdot \frac{1}{30\delta}) \approx \frac{3}{4} exp(-(\frac{C_{17}}{\sqrt{\delta/50}} \cdot \frac{1}{30\delta})^2) $. If $\delta$ is small enough, we have $C_{19}\leq C_{18}/2$. Notice that in fact $C_{19}$ can be made arbitrarily small in comparison to $C_{18}$, by choosing the constant $\delta$ small enough. Hence, by a union bound, the probability that ($\mathcal{E}_1$) does not happen or ($\mathcal{E}_2$) does not happen is at most $1-{C_{18}} + C_{19} \leq 1-{C_{18}} + C_{18}/2 = 1-{C_{18}}/2$. Thus, we conclude that there is a constant probability $C_{20} \geq C_{18}/2>0$ that events ($\mathcal{E}_1$) and ($\mathcal{E}_2$) both happen. In that case, the total number of light-blue and extra-light blue nodes, including their charges, exceeds $C_{17}\sqrt{n/\kappa}/2$. 

On the other hand, there is a constant probability that the light red nodes and the extra-light red nodes have a negative charge overall. In that case, the two populations have an additive difference of at least  $C_{17} \sqrt{n/\kappa}/2$. Recall that the numbers of clear blue and clear red are the same. If the light blue and extra-light blue exceed $C_{17} \sqrt{n/\kappa}/ 2$ and the light red and extra light red are non-positive, in relative terms compared to the populations $p_i$ and $p_j$ which are in  $[\frac{0.18}{\kappa}, \frac{1.5}{\kappa}]$, this is a gap of more than $g_{ij} \geq {C_1}/{\sqrt{n/\kappa}}$, if we choose the constant $C_{17}$ large enough as a function of the given $C_1$.
\end{proof}


\appendix
\section{Missing Details of the Analysis: Proving the Seven Properties}
\label{app:details}
We first prove properties (P1), (P4), (P6), and (P7), which are the easier ones. Then, we provide the proofs of properties (P2), (P3), and (P5), which will require more care and effort. At the end, we discuss how the adversary's effect can be incorporated into the analysis.
 
\subsection{Properties (P1), (P4), (P6) and (P7)}

\textbf{\Cref{lem:P1}} (\textbf{Property (P1)}) \emph{Suppose that $k=O(\sqrt{n/\log n})$, for a small enough constant in the $O$-notation. With high probability, super-weak opinions remain super-weak, and weak opinion remains weak.}
\medskip

\begin{proof} 
First, we argue that super-weak opinions remain super-weak. Consider a super-weak opinion $j$ so that $p_j\leq 1/(10k)$. We have $\mathbb{E}[p^{new}_j] = p_j (1 + p_j - \Sigma_2)$. In particular, we expect $n p^2_j$ nodes to move in to opinion $i$ while $n p_j \Sigma_2$ nodes move out. Suppose that $np^2_j \geq \log n$.
Then, using a standard Chernoff bound, we get that with high probability, the number of nodes that move in is no more than $2(np^2_j)$, and the number of nodes that move out is no less than $(np_j\Sigma_2)/2$. Notice that $\Sigma_2 = \sum_{i=1} p^2_i \geq 1/k$, because given the constraint $\sum_{i} p_i =1$, the summation $\sum_{i=1} p^2_i$ is minimized when the terms are equal. Hence, the out moves are still a $2.5$ factor larger than the in moves, which means $p_j$ can only decrease. Now suppose in the complementary case that $np^2_j \leq \log n$. Then, with high probability, the number of nodes that move in is no more than $5\log n$. Hence, even ignoring the out moves, at the end of the round, we have $p_j \leq \sqrt{\frac{\log n}{n}} + \frac{5\log n}{n} \leq 1/(10k)$, with high probability. This inequality uses the assumption that $k=O(\sqrt{n/\log n})$, for a suitably small constant in the $O$-notation.  

We now argue that weak opinions remain weak. Let $i$ be the plurality opinion and $j$ be a weak opinion. By definition of weak, we have $p_j \leq p_i/5$. We moreover have $\mathbb{E}[p^{new}_i] = p_i (1 + p_i - \Sigma_2)$ and $\mathbb{E}[p^{new}_j] = p_j (1 + p_j - \Sigma_2).$ In other words, the number of nodes that move in/out of $p_i$ and $p_j$ are respectively $np_i(p_i - \Sigma_2)$ and $np_j(p_j-\Sigma_2)$. Using the Chernoff-Hoeffding bound, we get that the additive deviation in the number of in/out moves from these expectations is at most $O(\sqrt{\frac{p^2_i \log n}{n}})$, with high probability. Hence, we can say that with high probability, $p^{new}_{i} \geq p_{i} (1 + p_i - \Sigma_2) -O(\sqrt{\frac{p^2_i \log n}{n}})$ and $p^{new}_{j} \leq p_{j} (1 + p_j - \Sigma_2)+ O(\sqrt{\frac{p^2_i \log n}{n}})$. Therefore, since $p_i\geq \frac{n}{k}$, we have 
\begin{eqnarray*}
\frac{p^{new}_i}{p^{new}_j} &\geq& \frac{p_i}{p_j}\cdot \frac{(1 + p_i - \Sigma_2)}{(1 + p_j - \Sigma_2)} \cdot \big(1-O(\sqrt{\frac{\log n}{n}})\big) \\
& 
 \geq& \frac{p_i}{p_j}\cdot (1 + \frac{p_i - p_j}{1+p_j - \Sigma_2}) \cdot  \big(1-O(\sqrt{\frac{\log n}{n}})\big) \\
 &\geq& \frac{p_i}{p_j}\cdot (1 + \frac{p_i}{4}) \cdot  \big(1-O(\sqrt{\frac{\log n}{n}})\big) 
 \geq \frac{p_i}{p_j} (1 + \frac{p_i}{8}) > 5.
\end{eqnarray*} 
Here, the penultimate inequality uses the fact that $p_i \geq 1/k$ and the assumption that $k=O(\sqrt{n/\log n})$ for a sufficiently small constant in the $O$-notation. Hence, opinion $j$ will have at least opinion $i$ which is a $5$ factor stronger, and thus, opinion $j$ will remain weak compared to the (potentially new) maximum opinion.
\end{proof}

\medskip
\noindent\textbf{\Cref{lem:P4} }(\textbf{Property (P4)}) \emph{For any two strong opinions $i$ and $j$ and any desirably large constant $C_5\geq 0$, with high probability, in $O(\log n)$ phases, we will have $\max\{g_{ij}, g_{ji}\} \geq \frac{C_5\sqrt{\log n}}{\sqrt{n/\kappa}}$.}
\medskip

\begin{proof} The proof is somewhat standard for explosive processes, and is similar to \cite[Lemma 8]{doerr2011stabilizing} to a great extent. We thus provide only a sketch. We call a phase \emph{successful} if one of the following two conditions holds: 
\begin{itemize}
\item [(1)] at the beginning of the phase, we have $\max\{g_{ij}, g_{ji}\} < \frac{C_1}{\sqrt{n/\kappa}}$ and at the end of the phase, we have $\max\{g^{new}_{ij}, g^{new}_{ji}\} \geq \frac{C_1}{\sqrt{n/\kappa}}$, 
\item [(2)] at the beginning of the phase, we have $\max\{g_{ij}, g_{ji}\} \geq \frac{C_1}{\sqrt{n/\kappa}}$ and at the end of the phase, we have $\max\{g^{new}_{ij}, g^{new}_{ji}\} \geq g_{ij}(1+\delta/10)$. 
\end{itemize}

By \Cref{lem:P2} the probability of failing in the first case is at most $1-C_2$. Moreover, by \Cref{lem:P3}, the probability of failing in the second case is at most $exp(-C_4 x^2)$ where $x = g_{ij} \cdot \sqrt{n/\kappa}$. 

We define a \emph{success streak} to be a sequence of consecutive successful phases until the first failure. Let $Y$ denote the random variable that is the length of bounded-length success streak, in terms of the number of phases. Note that a success streak may be unbounded, meaning that we never fail. For each finite $t\geq 1$, the probability of a $t$-length streak is $$\Pr[Y=t] \leq (1-C_2) \bigg(\prod_{\ell=1}^{t-1} \big(1- exp(- C_{16} \cdot (1+\delta/10)^\ell)\big)\bigg) \cdot exp(- C_{16} \cdot (1+\delta/10)^t)\big),$$
for some constant $C_{16}>0$. If a streak goes for more than $T=O(\log_{1+\delta/10} \sqrt{\log n})$ phases, then we reach a setting where $\max\{g_{ij}, g_{ji}\}  \geq \frac{C_5\sqrt{\log n}}{\sqrt{n/\kappa}}$. The probability of a streak going for more than $T$ phases is $$(1-C_2) \bigg(\prod_{\ell=1}^{T} \big(1- exp(- C_{16} \cdot (1+\delta/10)^\ell)\big)\bigg)> C_{17},
$$
for some constant $C_{17}>0$. Hence, if we check $\Theta(\log n)$ streaks, with high probability, one of them will be longer than $T$ phases, and will thus get us to $\max\{g_{ij}, g_{ji}\} \geq \frac{C_5\sqrt{\log n}}{\sqrt{n/\kappa}}$. Because of the distribution of $Y$, we see that the distribution of bounded-length streaks has an exponentially decaying tail as a function of the finite length $t\geq 1$. Thus, similar to \cite[Lemma 7]{doerr2011stabilizing}, one can see that the time needed for $\Theta(\log n)$ bounded-length streaks is $O(\log n)$ phases, with high probability. Thus, with high probability, after $O(\log n)$ phases, $\max\{g_{ij}, g_{ji}\} \geq \frac{C_5\sqrt{\log n}}{\sqrt{n/\kappa}}$.
\end{proof}

\medskip
\noindent\textbf{\Cref{lem:P6}} (\textbf{Property (P6)})\emph{
Consider the $i^{th}$ epoch where at most $\kappa=\lfloor k (5/6)^{i-1}\rfloor \geq 2$ not-super-weak opinions exist, at the beginning. W.h.p., within $O(\kappa \log n)$ rounds, the end-of-time arrives and we have $p_{max}=\max_{i} p_i \geq \frac{1.5}{\kappa}$.}
\medskip

\begin{proof} 

Suppose that in the first $O(\log n)$ phases, we have $p_{max}=\max_{i} p_i \leq \frac{1.5}{\kappa}$. Then, by applying properties (P4) and (P5), we get that within $O(\log n)$ phases, any two strong opinions $i$ and $j$ have a gap of at least $4$, that is, $\max\{g_{ij}, g_{ji}\} \geq 4$. That implies that the plurality opinion is a $5$ factor stronger than any other opinion. But since not-super-weak opinions must have at least $9/10 $ fraction of support in total, and there are only $\kappa$ of them, the plurality opinion must have support at least $\frac{5}{\kappa+4} \cdot \frac{9}{10} \geq \frac{1.5}{\kappa}$. In other words, we have reached the \emph{end-of-time} scenario.
\end{proof}

\medskip
\noindent\textbf{\Cref{lem:P7}} (\textbf{Property (P7)})\emph{Suppose that in the $i^{th}$ epoch, which starts with at most $\kappa = \lfloor k (5/6)^{i-1}\rfloor$ not-super-weak opinions, we have reached the end-of-time and have $\max_{i} p_i \geq \frac{1.5}{\kappa}$. Then, within $O(\kappa \log k)$ additional rounds, with high probability, at most $5\kappa/6$ not-super-weak opinions remain and thus, the next epoch begins.}\medskip

\begin{proof} Consider when the \emph{end-of-time} arrives. At most $5\kappa/6$ opinions can have support at least $\frac{1.2}{\kappa}$. Let $j$ be any other opinion such that $p_j < \frac{1.2}{\kappa}$. We now argue that opinion $j$ will become super-weak within $O(\kappa \log k)$ rounds. Notice that at any time during the next $O(\kappa \log k)$ rounds, we have $\max_{i} p_i \geq \frac{0.9}{\kappa}$, simply because there are at most $\kappa$ not-super-weak opinions and these must have at least $0.9$ fraction of the support in total. 

Let us examine the evolution of the ratio $\frac{p_{max}}{p_j}$. At the moment, we have $\frac{p_{max}}{p_j} \geq 1.2$. If we let $i$ to be the current plurality opinion, using a calculation similar to \Cref{lem:P1}, we get that with high probability, $\frac{p^{new}_i}{p^{new}_j} \geq \frac{p_i}{p_j} (1 + \frac{p_i}{20})$. Hence, even though the plurality opinion holder may switch, the ratio $\frac{p_{max}}{p_j}$ grows by a factor of $1+ \frac{p_{max}}{20} > 1+ \Omega(\frac{1}{\kappa})$.  Since $\frac{p_{max}}{p_j}$ starts above $1.2$, within $O(\kappa \log k)$ rounds, this ratio grows beyond $10k$. Thus, opinion $j$ is super-weak at that point.
%
%
\end{proof}

\subsection{Bringing Back the Adversary to The Analysis}
\label{app:adversary}
For any valid opinion, we will show that the impact of an $F$-bounded adversary, for $F=\eps\sqrt{n}/k^{1.5}$ where $\eps$ is a desirably small constant, is at most as strong as the random deviations that we took into account in our analysis. For non-valid opinions, which are introduced to the system by the adversary, we group all these non-valid opinions as one opinion $i=k+1$. Clearly, this can not increase our chances of converging to a valid opinion, it can only strengthen the non-valid opinions. Despite that, we show that this single non-valid opinion is super-weak and remains super-weak at all times.

First, in the case of the adversary's influence on valid opinion, this can only impact proofs of properties (P1), (P2), (P3), (P5), and (P7). This is because properties (P4) and (P6) do not deal with the dynamics directly, and merely rely on the other properties. Let us now examine these potentially impacted properties. 

In proof of property (P1), for each opinion, we took into account a deviation of size at most $O(\sqrt{n\log n/k^2})$. The adversary's influence is $F=\eps\sqrt{n}/k^{1.5} = o(\sqrt{n\log n/k^2})$. Hence, in this case, the adversary's impact is at most a second-order term in the deviations, and thus, since we anyways did not rely on the constant in the deviations, the proof goes through as before. The effect in property (P7) is similar, it is much smaller than the deviations that are already taken into account, and thus the analysis remains the same. Notice that in the case of (P1) and (P7), there is even some slack and we could have tolerated an even larger $F$. The bottleneck on that appears in properties (P2), (P3), (P5).

The adversary's effect in (P2), (P3), (P5) is similar. Recall that all three of these are about one phase, which is made of $\delta k$ rounds. During these rounds, the adversary can increase the number of nodes of one opinion by at most $\delta k F \leq \epsilon \delta \sqrt{n/ k}$. For instance, in the case of opinion $i$, we consider these as a small increase in the number of light blue nodes. As argued before, the number of extra-light blue nodes created because of these light blue nodes is at most $10\epsilon \delta^2 \sqrt{n/ k}$. Hence, overall, the effect of the adversary is increasing the number of light blue or extra light blue nodes by $\pm 2\epsilon \delta \sqrt{n/ k}$. In terms of fraction of nodes, this is at most ${2\epsilon \delta}/\sqrt{nk}$. Choosing $\eps$ a small enough constant, this is much smaller than the overall deviation $C_7x/\sqrt{\kappa n}$ taken into account in proving (P3), the $C_7\sqrt{\log n}/\sqrt{\kappa n}$ deviation taken into account in proving $C_7x/\sqrt{\kappa n}$, and the $C_{17}/\sqrt{\kappa n}$ taken into account in proving (P2). Hence, in all of these, the adversary's impact is far below the deviations that we took into account, and therefore, the analysis remains effectively the same.


\end{document}